\documentstyle[epsfig, twocolumn]{iso98}

\newcommand{\thepapername}{}

\newcommand{\um}{\,$\mu$m}
\newcommand{\cm}{\,cm}

\begin{document}

\setlength{\parindent}{0pt}
\setlength{\parskip}{ 10pt plus 1pt minus 1pt}
\setlength{\hoffset}{-1.5truecm}
\setlength{\textwidth}{ 17.1truecm }
\setlength{\columnsep}{1truecm }
\setlength{\columnseprule}{0pt}
\setlength{\headheight}{12pt}
\setlength{\headsep}{20pt}
\pagestyle{veniceheadings}

\title{\bf INFRARED EMISSION FROM RADIO-LOUD ACTIVE NUCLEI}

\author{{\bf Ilse~M.~van~Bemmel$^1$, Peter~D.~Barthel$^2$, Thijs~de~Graauw$^3$} \vspace{2mm} \\
$^1$European Southern Observatory, Garching bei M\"unchen, Germany \\
$^2$Kapteyn Astronomical Institute, Groningen, The Netherlands \\
$^3$Dutch Space Research Organisation, Groningen, The Netherlands}

\maketitle

\begin{abstract}

In order to test the unification scheme for double-lobed radio sources,
the far-infrared properties of matched samples of radio-galaxies and
radio-loud quasars were studied using ISOPHOT.  The quasar data were
complemented with nearly simultaneous submillimeter and centimeter radio
data.  The results show that quasars are generally brighter than
radio-galaxies in their far-infrared output and that beamed non-thermal
radiation must be excluded as source for this infrared excess.  However,
non-thermal flares or optically thick tori can still reconcile these
results with the unification scheme. 
 \vspace {5pt} \\
Key~words: active galactic nuclei; unification; spectral energy distributions

\end{abstract}

\section{INTRODUCTION}

The unified theory for radio-loud active galaxies (\cite{pdb89};
\cite{urry95}) predicts the existence of a dusty torus surrounding the
central engine.  Models for the reprocessing of the hard non-thermal nuclear
radiation by this torus show that at wavelengths longward of 60\um \ the
torus will be optically thin (\cite{pier92}; \cite{gran94}).  However,
in the IRAS observations quasars are two to four times brighter at 60\um
\ than radio galaxies (\cite{heck92}).  In some special geometries of
the torus, models can account for this infrared excess.  Also the torus
could still be optically thick at 60\um, or there could be a significant
amount of angle dependent non-thermal contamination in
quasars. Non-thermal beamed radiation has been proposed as a
possible solution (\cite{hes95}).

In order to discover the true cause of the infrared excess in quasars,
observations are needed at longer wavelengths.  All current torus models
predict that beyond 80\um \ the thermal emission will be isotropic. 
According to unified schemes, the output of quasars should then
equal that of matched radio-galaxies at these long far-infrared wavelengths. 
ISOPHOT observations of quasars and radio-galaxies have been obtained to
test this hypothesis.  To quantify any possible non-thermal
contamination in quasars, the latter were in addition observed
with SCUBA at the JCMT, and the NRAO Very Large Array (VLA).

\section{SAMPLE SELECTION AND OBSERVATIONS}

The target sample is selected using several criteria. First
of all the objects must have the edge-brightened, double-lobed radio structure
typical for Fanaroff \& Riley class II sources (\cite{fr78}). Six matched
pairs were constructed, containing one quasar and radio-galaxy each, 
having comparable redshift, radio-lobe power and angular size. 
Lobe radio power is a reasonably good measure of the active nucleus
power, and is furthermore independent of the orientation of the objects. 

Observations were carried out with ISOPHOT at three wavelengths, during
the period October 1996 to April 1998.  We made raster-mode mini-maps
with C100 and C200 (AOTs P3 and P22).  The rasters have a size of
$3\times3$ (C100) and $4\times2$ (C200) in Y$\times$Z direction.  The
filters used were the 60\um, 90\um \ (C100) and 160\um \ (C200).
Observations of only
four pairs were completed during the lifetime of ISO; for the other two
pairs only the radio-galaxy was observed.  For the quasars these data
were complemented with nearly simultaneous VLA A-array data at 6, 2, 1.2
and 0.7\cm \ (April 17, 1998) and JCMT SCUBA data at 2\,mm \ and 850\um
\ (May 28, 1997), addressing their non-thermal radio core strengths. 
These observations were made in order to detect possible non-thermal
contamination of the far-infrared emission.  Since the effect of
non-thermal emission in radio-galaxies is negligible (\cite{hoek98}),
these objects were not observed with the VLA and SCUBA. 

\begin{table*}[!ht]
\caption{\em Flux densities in mJy as plotted in Figure 1 (table added after
submission to proceedings). For 3C\,351 the flux density at 6\cm \ is
from our VLA observation, while in Figure 1 we plotted the datapoint
from Bridle et al. (1994).}

\begin{center}
\leavevmode
\footnotesize
\begin{tabular}[h]{lrrrrrrrrr}
\hline \\[-5pt]
Object & $F_{6cm}$ & $F_{2cm}$ & $F_{12mm}$ & $F_{7mm}$ & $F_{2mm}$ &
$F_{850\mu}$ & $F_{160\mu}$ & $F_{90\mu}$ & $F_{60\mu}$ \\
\hline \\[-5pt]
3C\,19 & -- & -- & -- & -- & -- & -- & $\leq$25 & $\leq$50 & $\leq$45 \\
3C\,42 & -- & -- & -- & -- & -- & -- & 21 $\pm$ 6 & $\leq$55 & $\leq$50 \\
3C\,67 & -- & -- & -- & -- & -- & -- & $\leq$40 & $\leq$80 & 72 $\pm$ 35 \\
3C\,277.1 & $\leq$15 & 22.7 $\pm$ 0.7 & 21.5 $\pm$ 1.0 & 9.6 $\pm$ 1.8 & 
$\leq$17 & -- & 31 $\pm$ 3 & 36 $\pm$ 14 & 55 $\pm$ 28 \\
3C\,323.1 & 38 $\pm$ 2 & 35 $\pm$ 1.3 & $\leq$8.5 & $\leq$10 & $\leq$14 & 
$\leq$12 & 14 $\pm$ 6 & 41 $\pm$ 20 & 109 $\pm$ 36 \\   
3C\,334 & 167.7 $\pm$ 0.8 & 112.3 $\pm$ 0.4 & 91.4 $\pm$ 0.7 & $\leq$10 & 
20 $\pm$ 10 & 15 $\pm$ 8 & 56 $\pm$ 8 & 89 $\pm$ 16 & 132 $\pm$ 33 \\
3C\,351 & 9.2 $\pm$ 1.2 & -- & $\leq$5.0 & $\leq$10 & -- & -- & 
133 $\pm$ 14 & 262 $\pm$ 18 & 323 $\pm$ 71 \\
3C\,460 & -- & -- & -- & -- & -- & -- & 73 $\pm$ 22 & $\leq$50 & $\leq$22 \\  
\hline \\
\end{tabular}
\end{center}
\end{table*}

\section{DATA REDUCTION}
\subsection{ISO Data Reduction}

All ISO data have been processed with OLP version 6.11 and raw data have
been reduced using the Phot Interactive Analysis tool (PIA) version
7.2(e).  The steps from ERD to AAP level were all done using the default
procedures.  Deglitching at ERD level was done with the two-treshold
method, setting sigmas of 3.0 (flagging) and 0.5 (reacceptance).  At SRD
level reset interval correction was applied, as well as dark current
subtraction per raster point.  Afterwards the data were deglitched again
with a sigma of 2.5 (source measurements) or 0.5 (FCS measurements).  At
SCP level no significant corrections were applied.  The signal of the
FCS was in most cases outside the hard extrapolation limits of PIA,
therefore it was decided to use only default responsivities for all
sources at all wavelengths. 

\subsection{ISO Flux Determination}

The final flux was determined using an IDL program called ana\_mini\_map,
developed by the PHOT support center in Heidelberg. This program uses
each pixel of the detector as a scan across the source. Thus no map is
made, instead it determines the background flux from the two off-source raster
positions immediately before and after the on-source raster position.
This off-source flux is compared to the on-source flux. If a signal is
present, all pixels will show a positive difference between on- and
off-source positions. However, due to the faint flux levels, noise
caused a lot of false results for some pixels. To avoid using noisy
data, an IDL tool was developed to determine the noise on the pixels used
by ana\_mini\_map. Pixels with extremely high noise on either one of
the off-source positions or the on-source position were excluded from
the flux determination. This produced a reliable flux for all the
detections (errors are generally less than 10  per~cent).

Tests were conducted by comparing the results to DIRBE background
measurements and IRAS detections of our sources.  Both DIRBE and IRAS
reproduced our results within a 10 per~cent error limit.  However, due
to the large calibration errors for faint sources, the fluxes presented
here are to be taken with errors of about 30 per~cent (\cite{lem96}). 
In order to present the results as clearly as possible, only the
statistical errors are given by the error bars in the plots, thus
providing better insight into the real meaning of the detection. 

\subsection{VLA And SCUBA Data Reduction}

The VLA data were reduced using the AIPS version of 15 April 1998. 
Calibration was done using the standard flux calibrators and nearby
phase calibrators.  After calibration the maps were cleaned using CLEAN
and the core flux density values were determined using the MAXFIT
procedure.  3C\,351 caused some problems at 6\cm \ and 2\cm, because of
the strong hotspots in this source.  The exact flux of the core in these
bands has not yet been determined. The 6\cm \ core flux of \cite{brid84}
was plotted instead, since no strong variations have been reported of
this source at cm wavelengths.  For all sources the errors are smaller
than the symbols in the plots. 

The SCUBA data were reduced using the SCUBA Data Reduction Facility
(SURF) and Kernel Application Package (KAPPA).  Not all sources were
observed at all bands.  Only sources with an observation have a SCUBA
upperlimit, and only 3C\,334 has been marginally detected
($\sim2\sigma$), be it in both bands.

\section{RESULTS}

The resulting broad band spectra are plotted for each pair in Figure 1. 
It is immediately clear that in most cases the quasar is still brighter
than the radio-galaxy, even up to the longest ISOPHOT wavelengths, with
only one exception.  The quasars are detected at all ISOPHOT bands,
whereas in only one out of four observations the radio-galaxy was
detected. From the
additional VLA and SCUBA data it is evident that the quasar cores do not
contribute significantly to the far-infrared emission observed with ISO,
since the sub-mm points and VLA 7\,mm data would have to be much higher
in that case. 

From the simple black-body fits to the ISO data it is also clear that
two of the four radio-galaxies need a lower temperature than the
quasars.  This could imply that in radio-galaxies we observe a cooler
dust component, which is in agreement with the unification theory,
assuming a dust distribution with a temperature gradient.  In one case
the black-body temperature could not be determined, because there is no
detection of the radio-galaxy at any wavelength (3C\,19).  In the final
case the far-infrared spectra of the quasar and radio-galaxy are
identical, which is surprisingly enough the pair with compact sources. 

\begin{figure*}[!ht]
  \begin{center}
    \leavevmode
  \centerline{\epsfig{file=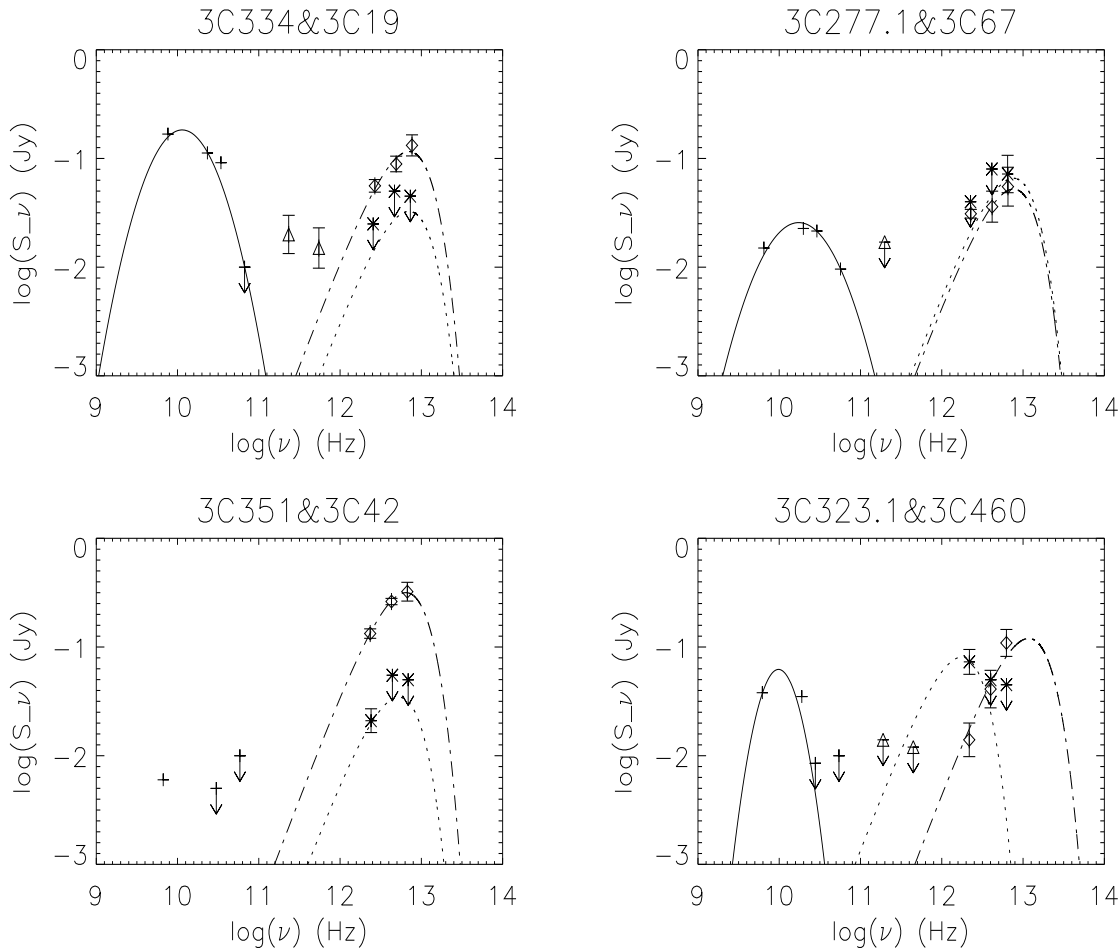,width=16cm,height=13cm}}
  \end{center}
  \caption{\em The complete spectra for the four pairs observed with
    ISO. For the quasars: plusses are VLA data, 
    triangles are SCUBA data, red diamonds are ISO data. The 
    solid line is an empirical fit to the radio-core spectra, the 
    dash-dotted line is a black body fit to the ISO data. For the 
    radio-galaxies: asterisks are ISO data and the dotted
    line is a black-body fit.}
\end{figure*}

\section{DISCUSSION}

\subsection{Non-thermal Contamination}

Radio-quiet and radio-loud quasars show remarkable similarities in their
broad band spectra.  The similar infrared to submillimeter slope and
far-infrared turnover frequency point to an identical source of infrared
emission.  Since radio-quiet quasars have no significant non-thermal
component, the infrared emission must be thermal (\cite{san89}). 
Assuming the models for thermal torus emission show the correct trends,
an aspect angle dependent non-thermal contamination must be present in
radio-loud quasars to reconcile the observed infrared excess in these
objects with the unification theory.  There are two possibilities:
continuous beaming, or flaring, which requires a
rapidly varying beamed component. 

Continuous beaming is in most cases not more than a 30 per~cent effect
(\cite{hoek98}) and has been solidly ruled out for causing the excess in
at least two quasars (\cite{ivb98}). Here, only one quasar is detected at
7\,mm and the detections with SCUBA are marginal at best. In our results
continuous non-thermal contamination of the far-infrared is
negligible, as can also be seen from the fits to the radio data in Figure~1.  
It is important to use only core flux densities, as this is the only
source of beamed non-thermal radiation in active galaxies. When using
the total radio flux densities as done by Haas et al. (1998), the
centimeter flux densities will be dominated by emission from the radio
lobes. These have generally steep spectra and are not beamed, thus
they do not contribute significantly to the infrared emission in these sources.

A varying component can easily be missed.  Strongly beamed radiation
from non-thermal flares has been observed in blazars (\cite{brwn89});
such flares are both very strong and short-lived, with typical
timescales of several days.  They could be observed in quasars due to
the small aspect angle of these objects.  In previous work two out of
three quasars showed high 3\,mm fluxes, which could not be fitted with a
radio core model and were suggestive of flaring (\cite{ivb98}).  In
radio-galaxies beamed flare radiation is much less likely to be
observed, because of the
larger aspect angles in these sources.  Strong flares can cause large
differences in far-infrared output, whilst being missed in the
non-thermal core spectra. 

\subsection{Optically Thick Torus}

The persistence of the infrared-excess at longer wavelengths can also be
explained if the torus is optically thick up to 140\um \ at least. 
Arguing in favour of this concept is the fact that a lower temperature
is needed for the black-body fit in at least two of the radio-galaxies. 
Here maybe only the cooler edge of the torus is seen, while in quasars the hot
inner parts dominate the infra-red output.  This is in contradiction
with the present models for dusty tori, but in broad agreement with
predictions from the unified models.  Since changing the models for the
torus also has implications for the X-ray spectra observed in these
sources, new modelling has to be done to study this possibility in more
detail. 

\subsection{Evolution Between Types}

Evolution has been proposed as a solution to the infrared-excess
(\cite{ivb98}).  This would then imply that quasars are younger and
still contain more dust in their host galaxy, whereas radio-galaxies are
older and have processed their dust and gas into stars.  A link with the
ultra-luminous galaxies has been made in this respect as these being
very young quasars who are in the phase of birth (\cite{san88}). 
However, during a recent conference on this topic, most evidence showed
that these so-called ULIRGs are in many cases driven by strong
starbursts and not by a central active nucleus (Monsters or Babies:
What powers ultra-luminous infrared galaxies?, Ringberg Castle, 1998).
The fact that we find one pair in which the far-infrared spectra are
identical, also argues against evolution between the types. Any
evolutionary model is obviously in conflict with orientation based 
unification concepts.

\section{CONCLUSIONS}

Observations of matched pairs of quasars and radio-galaxies show that
the previously reported infrared-excess is still present in quasars, up
to 140\um \ restframe wavelength.  This cannot be accounted for by 
current torus models, without causing a contradiction with unified
schemes.  Continuous beaming has to be excluded in our objects, but
non-thermal flares can explain the excess, without necessary
adjustments to either theory.  Also the models for the torus have to
be examined in more detail to explain observed temperature differences
between quasars and radio-galaxies.  This is currently under
investigation and will be described in future publications. 

Future work will also include the search for flares in 3C\,334 with
SCUBA and the investigation of orientation dependence of optical
emission lines in a sample of double-lobed active galactic nuclei. 

\section*{ACKNOWLEDGEMENTS}

Our thanks to all people who helped us during the hard reduction
process, especially Martin Haas who provided a lot of useful tips and
advice.  Also thanks to Ronald Hes for initial involvement.  And thanks
to Eric Hooper, Mari Poletta, Belinda Wilkes, Thierry Courvoisier and
Rolf Chini for advice and discussions.  Thanks to Xander Tielens for
useful comments and advice while making this manuscript. 

ISO is an ESA project with instruments funded by ESA Member States
(especially the PI countries: France, Germany, the Netherlands and the
United Kingdom) and with the participation of ISAS and NASA.  PIA is a
joint development by the ESA Astrophysics division and the ISOPHOT
consortium.  The National Radio Astronomy Observatory is a facility of
the National Science Foundation operated under cooperative agreement by
Associated Universities, Inc.  The James Clerk Maxwell Telescope is
operated by The Joint Astronomy Centre on behalf of the Particle Physics
and Astronomy Research Council of the United Kingdom, the Netherlands
Organisation for Scientific Research, and the National Research Council
of Canada. 

\section{BIBLIOGRAPHIC REFERENCES}

\end{document}